\pdfoutput=1
\documentclass[12pt]{article}
\usepackage{graphicx}
\usepackage{amssymb,amsfonts,amsmath}
\usepackage{epsfig}
\def\lsim{\mathrel{\rlap{\lower3pt\hbox{\hskip1pt$\sim$}}
     \raise1pt\hbox{$<$}}}
\def\gsim{\mathrel{\rlap{\lower3pt\hbox{\hskip1pt$\sim$}}
     \raise1pt\hbox{$>$}}}
\newcommand{\beq}{\begin{equation}}
\newcommand{\eeq}{\end{equation}}
\newcommand{\bea}{\begin{eqnarray}}
\newcommand{\eea}{\end{eqnarray}}

\begin{document}
\title{Are there hadronic bound states above the QCD transition temperature?}
\author{Claudia Ratti$^a$, Rene Bellwied$^b$, Marco Cristoforetti$^{c,d}$, Maria Barbaro$^a$ \\
$^a$ \small{\it Universit\`a degli Studi di Torino and INFN, Sezione di Torino,}\\
\small{\it via Giuria 1, I-10125 Torino (Italy),}\\
$^b$ \small{\it University of Houston, Houston, TX 77204, USA,}\\
$^c$ \small{\it ECT*, strada delle Tabarelle 286, I-38123 Villazzano (Trento), Italy,}\\
$^d$ \small{\it LISC, Via Sommarive 18, Povo (Trento), I-38123  Italy.}}

\maketitle
\begin{abstract}
Recent lattice QCD calculations, at physical pion masses and small lattice
spacings that approach the continuum limit, have revealed that
non-diagonal quark correlators above the critical temperature are finite
up to about 2 $T_c$. Since the transition from hadronic to free
partonic degrees of freedom is merely an analytic cross-over, it is likely
that,
in the temperature regime between 1-2 $T_c$, quark and gluon quasiparticles
and pre-hadronic bound states can coexist. The correlator values, in
comparison to PNJL model calculations beyond mean-field, indicate that at
least part of the mixed phase resides in color-neutral bound states.
A similar effect was postulated for the in-medium
fragmentation process, i.e. for partons which do not thermalize with the
system and thus constitute the non-equilibrium component of the particle
emission spectrum from a deconfined plasma phase. Here, for the first time
we investigate the likelihood of forming bound states also in the
equilibrated, parton dominated phase above $T_c$ which is described by
lattice QCD.
\end{abstract}
\newpage
\section{Introduction}
The study of the QCD phase diagram and thermodynamics is receiving increasing attention in recent years. This field of physics is particularly appealing because
the deconfined phase of QCD can be produced in the laboratory, in the
ultrarelativistic heavy ion collision experiments at CERN SPS, BNL RHIC, CERN LHC and the future facilities
at the GSI (FAIR) and in Dubna (NICA). On the other hand, lattice calculations on QCD thermodynamics are reaching unprecedented levels of accuracy, with
simulations at the physical quark masses and several values of the lattice spacing approaching the continuum limit. In addition, the interpretation of lattice results in terms of phenomenological models is of fundamental importance in order to understand the microscopic structure of the QCD deconfined medium. The information that can be obtained from these complementary approaches will shed
light on the features of QCD matter under extreme conditions, one of the major
challenges of the physics of strong interactions.

One fundamental question is the nature of the effective degrees of freedom in the temperature range 1-2 $T_c$: the experimental results available so far show that the hot QCD matter produced
in the laboratory exhibits robust collective flow phenomena, which are well and
consistently described by relativistic hydrodynamics \cite{Teaney:2000cw,Teaney:2001av,Kolb:2003dz}.
The data are actually consistent with a near zero shear viscosity over entropy ($\eta$/s) ratio, which signals the
existence of a strongly interacting perfect fluid rather than a weakly coupled plasma state. In order to explain such a strong coupling between the degrees of freedom one either has to allow a strong enhancement in multi-parton interactions \cite{Xu:2004mz} or a modification of partonic states, i.e. quasi-particles or color-neutral bound states. 
The quasi-particle model has been recently compared to the latest lattice QCD data
\cite{Plumari:2011mk} resulting in a large, temperature dependent, effective mass for quarks and
gluons near $T_c$ \cite{Gorenstein:1995vm,LH1998,PC05}. However, the model cannot simultaneously reproduce bulk thermodynamics (pressure, energy density, interaction measure) and quark number susceptibilities \cite{Plumari:2011mk,susce0}.
Alternatively, the interaction between partons might be enhanced due to the existence
of a large number of binary bound states, mostly colored, in the QGP \cite{Shuryak:2004tx}. In addition, in-medium hadronization and the formation of color-neutral objects inside the partonic fireball, due to short formation times, have been postulated in Refs. \cite{Markert:2008jc,Bellwied:2010pr} for the non-equilibrium component of the particle emission spectrum from a deconfined plasma phase. In this paper we address the possible presence of bound states also in the equilibrated component of the QGP phase.

Correlations between non-diagonal quark flavors or between strangeness and baryon number have been proposed as a diagnostics to understand the nature of QCD matter immediately above the phase transition \cite{Koch:2005vg}. These observables can be calculated from first principles in lattice QCD, and several results have been published over the last few years \cite{susceptibilities,eos}. However, the rather large quark masses and lattice spacings used in these simulations did not allow to draw a decisive conclusion so far. Very recently, new lattice results have become available for these observables, with simulations at the physical quark masses and finer lattice spacings approaching the continuum limit \cite{Mukherjee:2011td,Borsanyi:}. These results, which are detailed in the next section, seem to favor a scenario in which bound states are present in the deconfined medium for a certain temperature range above $T_c$.

In this paper we are comparing the latest lattice QCD calculations of susceptibilities with the predictions of the Polyakov loop-extended Nambu Jona-Lasinio model \cite{Ratti:06}; the model correctly describes the chiral phase transition of QCD and incorporates aspects of the deconfinement phase transition through the coupling with the Polyakov loop.
Our purpose is to understand the nature of the relevant degrees of freedom; we concentrate our analysis in the temperature region just above the phase transition. By going beyond the mean-field approximation and using the Monte Carlo method applied to the PNJL model~ \cite{Cristoforetti:2010sn}, we incorporate fluctuations of the condensates (chiral, pion and kaon condensates) and of the Polyakov loop. In this way, we estimate quark number susceptibilities and baryon-strangeness correlations for a partonic quark-gluon plasma containing mesonic zero-modes and an additional degree of freedom (the Polyakov loop) which couples the different flavors. We attribute the difference with respect to lattice results to the presence of finite-momentum bound states in the QGP.

\section{Lattice QCD results}

\begin{figure}
\begin{minipage}{.48\textwidth}
%\hspace{.8cm}
\parbox{6cm}{
\scalebox{.65}{
\includegraphics{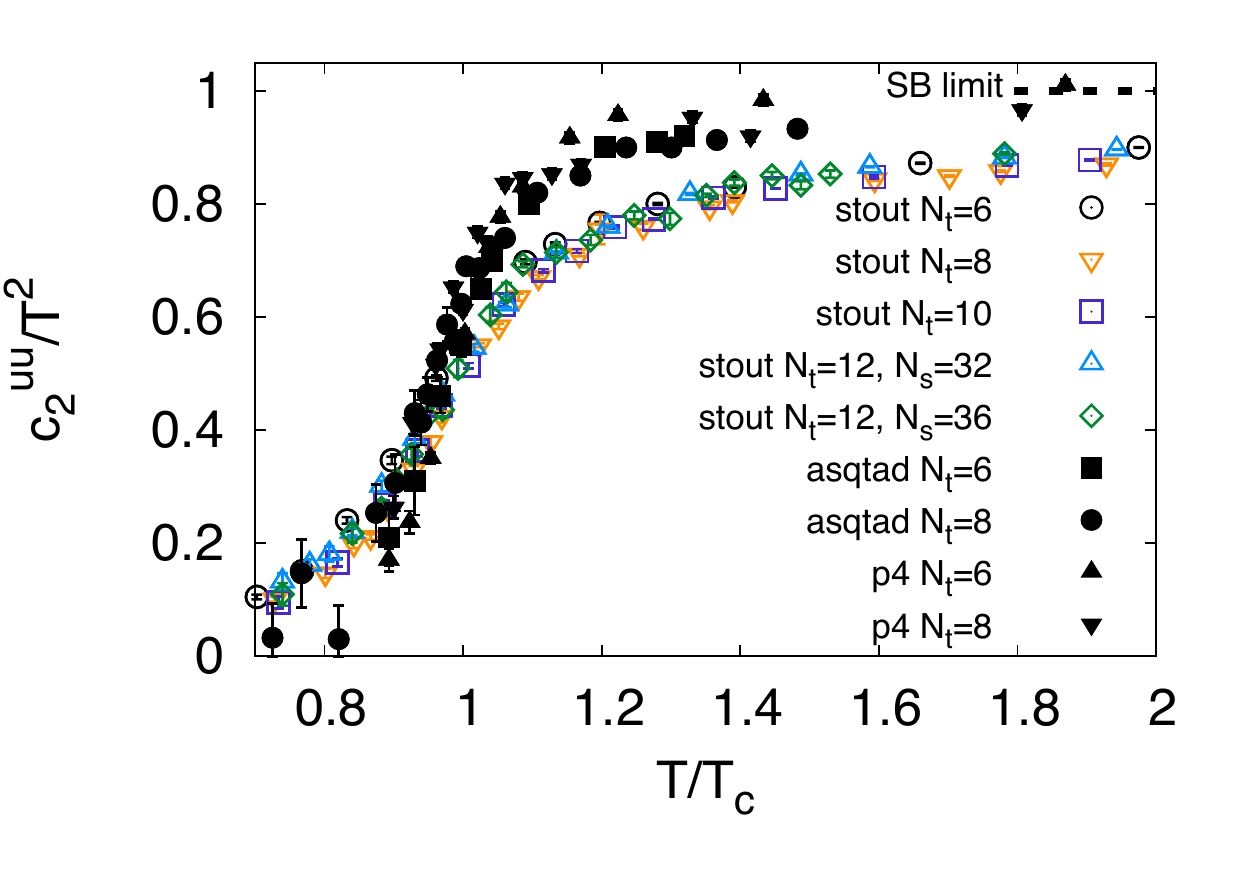}\\}}
\end{minipage}
\begin{minipage}{.48\textwidth}
%\hspace{.8cm}
\parbox{6cm}{
\scalebox{.65}{
\includegraphics{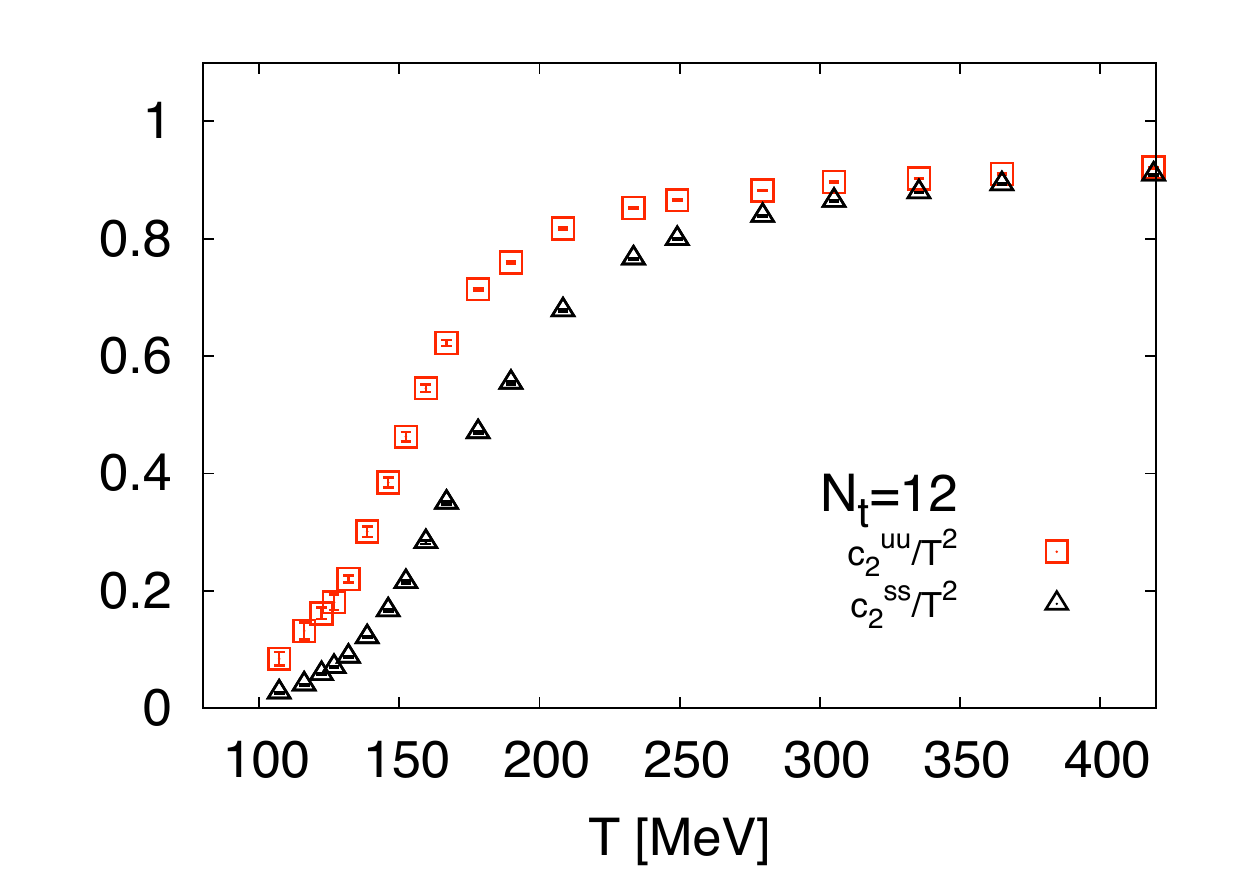}\\}}
\end{minipage}
%\centerline{(a)}
\caption{Left: Comparison between the lattice results for light quark number susceptibilities obtained with the stout \cite{Borsanyi:}, asqtad and p4 actions \cite{eos} (notice that on the horizontal axis the temperature is normalized by the different $T_c$ values obtained with the different actions). Right: comparison between the lattice results for light and strange quark number susceptibilities, obtained with the stout action at physical quark masses and $N_t=12$ (from Ref. \cite{Borsanyi:}). For the definition of $N_s$ and $N_t$ see eq. (\ref{vol}).}
\label{fig1}
%\vspace{-.4cm}
\end{figure}

Recent lattice calculations unambiguously show that the transition from the hadronic to the partonic system
at zero baryo-chemical potential is an analytic cross-over \cite{Aoki:2006we}. In such cases the critical, or transition, temperature is determined by the inflection point in the temperature dependence of the relevant observables. The main quantities that are used to determine
the transition temperature are the Polyakov loop, energy density and quark number susceptibilities for the deconfinement phase transition, and the quark condensates for the chiral phase transition.
The smooth behavior of all these QCD observables as functions of the temperature, which is evident in the most recent calculations that employ smaller lattice spacings and realistic quark masses, leads to interesting cross-over phenomena \cite{Borsanyi:2010bp}. For example, in the left panel of Fig. \ref{fig1} we show a comparison between the previously available lattice results for the light quark number susceptibilities \cite{eos} and the new Wuppertal-Budapest results obtained with physical quark masses and smaller lattice spacings \cite{Borsanyi:}: it is evident that, for the most recent data, the transition is less steep.

Both deconfinement, as expressed through the renormalized Polyakov loop or quark number susceptibilities, and chiral symmetry restoration, shown in the chiral condensates, experience therefore an extended transition region before reaching the fully deconfined and chirally symmetric state. At any given temperature in this cross-over range these parameters could thus be interpreted as signalling a mixed phase of degrees of freedom where bound states or chirally broken states will co-exist with free quarks and gluons according to the relative values of the Polyakov loop or the chiral condensates. This transition region was called ``semi-QGP" in Ref. \cite{Hidaka:2008dr}, as opposed to the ``full-QGP" at large temperatures, where the Polyakov loop is close to one and flat.

Furthermore, the latest lattice results signal a flavor dependence of quark number susceptibilities even in the light quark sector. The rise of the strange quark susceptibility with temperature is slower and takes place at larger temperatures  compared to the $u$ case, as shown in the right panel of Fig. \ref{fig1}. This feature was less pronounced in previous lattice results \cite{eos}, since in that case $m_s/m_{u,d}$ = 10, whereas the new results use a physical quark mass ratio ($m_s/m_{u,d}$ = 28.15).

This flavor difference between quark number susceptibilities in the light sector likely indicates that strange quarks experience deconfinement at slightly larger temperatures, compared to light quarks, thus implying a survival of strangeness-carrying hadrons in the QGP immediately above $T_c$.

%\begin{center}
%\end{center}

Another piece of evidence pointing in this direction is the behavior of the baryon-strangeness correlator $C_{BS}$ or the non-diagonal quark number susceptibilities.  Non-diagonal $u-s$ susceptibilities (shown in Fig. \ref{fig2}) exhibit a pronounced peak in the vicinity of the phase transition, and remain finite for relatively large temperatures above $T_c$. Similarly, the baryon-strangeness correlator (also shown in Fig. \ref{fig2}) only reaches the predicted value for a purely partonic QGP (approaching the Stefan-Boltzman limit) near 2 $T_c$. Although it was shown in Ref. \cite{Blaizot:2001vr} that correlations between different flavors are nonzero in perturbative QCD at large temperatures due to the presence of flavor-mixing diagrams, the lattice data exhibit a strong enhancement of these correlations in the vicinity of $T_c$, which survives up to relatively large temperatures above the transition \cite{Borsanyi:} and which cannot be accounted for by the perturbative QCD contribution alone.
Taking into account this behavior, one could come to the conclusion that in the region 1-2 $T_c$ the probability of forming color neutral bound states is quantifiable even in
the case of a fully equilibrated system of quarks and gluons as simulated through lattice QCD.

\section{PNJL model}
The 2+1-flavor PNJL model is specified by the Euclidean
action
\beq
{\cal S}_E(\psi, \psi^\dagger, \phi)= \int _0^{\beta=1/T} d\tau\int d^3x \left[\psi^\dagger\,\partial_\tau\,\psi + {\cal H}(\psi, \psi^\dagger, \phi)\right] + \delta{\cal S}_E(\phi,T)
\label{action}
\eeq
with the fermionic Hamiltonian density \footnote{$\vec{\alpha} = \gamma_0\,\vec{\gamma}$ and $\gamma_4 = i\gamma_0$ in terms of the standard Dirac $\gamma$ matrices.}:
\beq
{\cal H} = -i\psi^\dagger\,(\vec{\alpha}\cdot \vec{\nabla}+\gamma_4\,m_0 -\phi)\,\psi + {\cal V}(\psi, \psi^\dagger)~,
\label{H}
\eeq
where $\psi$ is the $N_f=3$ quark field and $m_0 = diag(m_u,m_d, m_s)$ is the quark mass matrix, with $m_u=m_d\neq m_s$. Quarks move in a background color gauge field $\phi \equiv A_4 = iA_0$, where $A_0 = \delta_{\mu 0}\,g{\cal A}^\mu_a\,t^a$ with the $SU(3)_c$ gauge fields ${\cal A}^\mu_a$ and the generators $t^a = \lambda^a/2$. The matrix valued, constant field $\phi$ relates to the (traced) Polyakov loop $\Phi$ as follows:
\beq
\Phi=\frac{1}{N_c}\mathrm{Tr}\left[\mathcal{P}\exp\left(i\int_{0}^{\beta}
d\tau A_4\right)\right]=\frac{1}{ 3}\mathrm{Tr}\,e^{i\phi/T}~.
\eeq
In a convenient gauge (the so-called Polyakov gauge), the matrix $\phi$ is given a diagonal representation
\beq
\phi = \phi_3\,\lambda_3 +  \phi_8\,\lambda_8~,
\eeq
which leaves only two independent variables, $\phi_3$ and $\phi_8$. The piece $\delta{\cal S}_E = -\frac{V}{ T}{\cal U}$ of the action (\ref{action}) controls the thermodynamics of the Polyakov loop. It will be specified later in terms of the effective potential, ${\cal U}(\Phi,T)$, determined such that the thermodynamics of pure gauge lattice QCD is reproduced for $T$ up to about twice the critical temperature for deconfinement.
At much higher temperatures where transverse gluons begin to dominate, the PNJL model is not supposed to be applicable.

The interaction ${\cal V}$ in Eq.~(\ref{H}) is defined as follows:
\bea
\mathcal{V}&=& -\frac
{G}{2}\sum_{f=u,d,s}\left[\left(\bar{\psi}_f\psi_f\right)^2+\left(\bar{\psi}_fi\gamma_5
\vec{\tau}\psi_f
\right)^2\right]
\nonumber\\
&+&\frac K2\left[\det_{i,j}\left(\bar{\psi}_i(1+\gamma_5)\psi_j\right)+
\det_{i,j}\left(\bar{\psi}_i(1-\gamma_5)\psi_j\right)\right].
\label{V}
\eea
Spontaneous chiral symmetry breaking is driven by the first term in Eq. (\ref{V}), while the second term breaks the axial $U(1)_A$ symmetry explicitly.
The NJL part of the model involves five parameters: the quark masses which we take equal for $u$- and $d$-quarks and heavier for $s$ quarks, the coupling strengths $G$ and $K$ and a three-momentum cutoff $\Lambda$. We take those from Ref. \cite{Buballa:2003qv}, they are listed in Table \ref{t1}.
\begin{table}
\begin{tabular}{c}
Parameters\\
\vspace{.3cm}
\begin{tabular}{|c|c|}
\hline
$\Lambda$ [GeV]&0.6023\\
\hline
$G\Lambda^2$&3.67\\
\hline
$K\Lambda^5$&24.72\\
\hline
$m_{0u,d}$ [MeV]&5.5\\
\hline
$m_{0s}$ [MeV]&140.7\\
\hline
\end{tabular}
\vspace{-.3cm}\\
\end{tabular}
\hspace{2cm}
\begin{tabular}{c}
Physical quantities\\
\vspace{.3cm}
\begin{tabular}{|c|c|}
\hline
$f_\pi$ [MeV]&92.4\\
\hline
$|\langle\bar{\psi}\psi\rangle_{u,d}|^{1/3}$ [MeV]&241.9\\
\hline
$|\langle\bar{\psi}\psi\rangle_{s}|^{1/3}$ [MeV]&257.7\\
\hline
$m_\pi$ [MeV]&139.3\\
\hline
$m_K$ [MeV]&497.7\\
\hline
%$G_V$&$G_S/8$\\
%\hline
%&\\
\end{tabular}
\vspace{-.3cm}\\
\end{tabular}
\caption{Left: NJL model parameters. Right: physical quantities used to fix the parameters.}
\label{t1}
\end{table}
The effective potential $\mathcal{U}(\Phi,T)$ which controls the dynamics of the Polyakov loop has the following  form:
\beq
{\cal U}(\Phi,T )=-\frac{1}{2}a(T)\,\Phi^*\Phi
+b(T)\,\ln\left[1-6\,\Phi^*\Phi+4\left({\Phi^*}^3+\Phi^3\right)
-3\left(\Phi^*\Phi\right)^2\right]
\label{u1}
\eeq
with
\beq
a(T)=a_0+a_1\left(\frac{T_0}{T}\right)
+a_2\left(\frac{T_0}{T}\right)^2,~~~~~~b(T)=b_3\left(\frac{T_0}{T}
\right)^3.
\label{u2}
\eeq
The parameters are taken from the literature:
\bea
a_0 = 3.51~,~~a_1 = -2.47~,~~a_2 = 15.22~,~~b_3 = -1.75~.
\nonumber
\eea
The critical temperature $T_0$ for deconfinement in the pure gauge sector is
fixed at 270 MeV in agreement with lattice results.

In the model, quarks acquire a constituent mass through their interaction with the chiral condensate. Due to the flavor-mixing term in the Lagrangian, the mass of a given flavor also gets contributions from the chiral condensates of the other quark flavors:
\bea
m_{i}=m_{0i}-\langle\sigma_i\rangle-\frac{K}{4G^2}\langle\sigma_j\rangle
\langle\sigma_k
\rangle=m_{0i}-2G\langle\bar{\psi_i}\psi_i\rangle+K\langle\bar{\psi_j}\psi_j
\rangle\langle\bar{\psi_k}\psi_k\rangle~~~~~~i\neq j\neq k.
\nonumber
\eea

In the PNJL model the partition function in momentum space is written as
\bea
\label{eq:pfuncvi}
	\mathcal{Z}&=&\int\mathcal{D}\phi\mathcal{D}\sigma_\alpha\,\mathcal{D}\pi_\alpha\mathcal{D} S_\alpha\,\mathcal{D} P_\alpha\exp\Bigg[\frac{V}{T}\Bigg(\frac{1}{2} T \sum_n\int\frac{\textrm{d}^3p}{(2\pi)^3}\ln\det[S^{-1}(i\omega_n,\vec{p},\mu_q)]
	\nonumber\\
	&&\hspace{5cm}-\mathcal{U}(\phi,T)-\mathcal{S}(\sigma_\alpha,\pi_\alpha,S_\alpha,P_\alpha)\Bigg)\Bigg]
\eea
where $\omega_n=(2n+1)\pi T$ are the Matsubara frequencies, $\sigma_\alpha$ and $\pi_\alpha$ are $18$ bosonic fields corresponding to the possible scalar and pseudoscalar condensates while $S_\alpha$ and $P_\alpha$ are additional $18$ auxiliary fields necessary in order to deal with the six-fermion interaction induced by the 't Hooft term. For symmetry reason, clearly $\langle\pi_\alpha\rangle=0$ at the mean field level, but we will let these fields fluctuate in the Monte Carlo approach, too. This will allow to take the contribution of mesonic zero-modes into account. $\mathcal{U}(\phi,T)$ is the Polyakov loop potential given above and the $\mathcal{S}(\sigma_\alpha,\pi_\alpha,S_\alpha,P_\alpha)$ is the bosonic action
\bea
 	\mathcal{S}_\text{E}^\text{bos}&=&-\left[\sigma_\alpha S_\alpha+\pi_\alpha P_\alpha+\dfrac{G}{2}\left[S_\alpha S_\alpha+P_\alpha P_\alpha\right]+\right.
	\nonumber\\
	&&\left.+\dfrac{K}{4}\mathcal{A}_{\alpha\beta\gamma}\left[S_\alpha S_\beta S_\gamma-3 S_\alpha P_\beta P_\gamma\right]\right],
\eea
where the constants $\mathcal{A}_{\alpha\beta\gamma}$ are expressed in terms of the Gell-Mann matrices according to
\begin{equation*}	 \mathcal{A}_{\alpha\beta\gamma}:=\dfrac{1}{3!}\varepsilon_{ijk}\varepsilon_{mn\ell}\left(\lambda_\alpha\right)_{im}\left(\lambda_\beta\right)_{jn}\left(\lambda_\gamma\right)_{kl}\qquad\text{for $\alpha,\beta,\gamma\in\{0,\dots,8\}$}.
\end{equation*}

In order to perform the integration in $S_\alpha$ and $P_\alpha$, the stationary phase approximation is used, choosing the fields $S_\alpha$ and $P_\alpha$ so as to minimize the integrand in the bosonized partition function. The necessary condition imposed on the fields is
\begin{equation}\label{spa}
	\begin{aligned}
	 	\sigma_\alpha+G S_\alpha+\frac{3 K}{4}\mathcal{A}_{\alpha\beta\gamma}\left[ S_\beta  S_\gamma- P_\beta  P_\gamma\right]&=0\,,\\
		\pi_\alpha+G P_\alpha-\dfrac{3 K}{2} \mathcal{A}_{\alpha\beta\gamma} S_\beta P_\gamma&=0\,.
	\end{aligned}
\end{equation}

The presence of a volume factor $V$ in the exponent of Eq.~\ref{eq:pfuncvi} makes it possible to compute the full partition function using Monte-Carlo techniques. In this way we consider not only the saddle-point contributions, but also configurations that correspond to fluctuations around the minima of the action.

The size of the volume is now specified according to the conventions adopted in lattice calculations. For a fixed extension of the lattice in the Euclidean time direction, the temperature is set by the lattice spacing $a$, and the volume size is related to the temperature:
\bea
	a=\frac{1}{N_t T}& \rightarrow &V=N_s^3 a^3=\frac{N_s^3}{N_t^3 T^3},
	\label{vol}
\eea
where $N_t$ is the number of lattice sites in the Euclidean time direction, and $N_s$ is the number of lattice sites in the space direction.
It follows that
\bea\label{eq:vk}
	V = k/T^3.
\eea
In particular here we used the typical values of $N_s/N_t$ used in Lattice simulations which correspond to $k=27$ and $k=64$.

Let us see now the definition of quark number susceptibilities. The thermodynamic potential can be expanded in a Taylor series in $\mu_q/T$ around zero chemical potential,
\bea
	 \Omega(T,\mu)=\frac{1}{VT^3}\ln\mathcal{Z}=\sum_{i,j=0}^{\infty}c_{ij}^{mn}(T)\left(\frac{\mu_m}{T}\right)^i\left(\frac{\mu_n}{T}\right)^j,
\eea
with
\bea
	 c_{ij}^{mn}(T)=\frac{1}{i!j!}\left.\frac{\partial^{i+j}\Omega}{\partial(\mu_m/T)^i\partial(\mu_n/T)^j}\right|_{\mu_m=\mu_n=0},
\eea
where only even terms survive due to $CP$ symmetry. The coefficients $c_{ij}^{mn}(T)$ are evaluated at $\mu_q=0$.
The baryon-strangeness correlator is defined in the following way:
\bea
C_{BS}=1+\frac{c_{2}^{us}+c_{2}^{ds}}{c_{2}^{ss}}.
\eea

Looking at the definition of the partition function we immediately see that the susceptibilities $c_{ij}^{mn}$ involve derivatives of $\ln\det[S^{-1}(i\omega_n,\vec{p},\mu_q)]$, which is the only term in $\mathcal{Z}$ which explicitly depends on the chemical potentials.

For the first derivative of the log of the partition function we have:
\bea\label{eq:c2uda}
	&&\frac{\partial\ln\mathcal{Z}(T,\mu_u,\mu_d,\mu_s)}{\partial\mu_q}=\nonumber\\
	 &&=\frac{\partial}{\partial\mu_q}\ln\int\mathcal{D}\sigma_\alpha\mathcal{D}\pi_\alpha\mathcal{D}\phi\exp\Big[\frac{V}{T}\ln\det S^{-1}(\mu_u,\mu_d,\mu_s)-\mathcal{S}_g[\phi]\Big]\nonumber\\
	 &&=\frac{1}{\mathcal{Z}(\mu_u,\mu_d,\mu_s)}\frac{V}{T}\int\mathcal{D}\sigma_\alpha\mathcal{D}\pi_\alpha\mathcal{D}\phi\frac{\partial\ln\det S^{-1}(\mu_u,\mu_d,\mu_s)}{\partial\mu_q}e^{-\mathcal{S}[\mu_u,\mu_d,\mu_s]}\nonumber\\
	&&=\frac{V}{T}\Big\langle\frac{\partial\ln\det S^{-1}(\mu_u,\mu_d,\mu_s)}{\partial\mu_q}\Big\rangle.
\eea
Going on in the same way for the second derivative, we obtain the coefficients $c_{11}^{q_1q_2}$ (quark number susceptibilities)
\bea\label{eq:c2ud}
	&&c_{11}^{q_1q_2}= \frac{1}{VT}\frac{\partial^2}{\partial\mu_{q_1}\partial\mu_{q_2}}\ln\mathcal{Z}= \frac{T^2}{VT^3}\Bigg(\frac{V}{T}\left\langle\frac{\partial^2}{\partial\mu_{q_1}\partial\mu_{q_2}}\ln\det S^{-1}(\mu_u,\mu_d,\mu_s)\right\rangle \nonumber\\
	&&+\left(\frac{V}{T}\right)^2\left\langle\left(\frac{\partial}{\partial\mu_{q_1}}\ln\det S^{-1}(\mu_u,\mu_d,\mu_s)\right)\left(\frac{\partial}{\partial\mu_{q_2}}\ln\det S^{-1}(\mu_u,\mu_d,\mu_s)\right)\right\rangle\nonumber\\
	&&-\left(\frac{V}{T}\right)^2\left\langle\frac{\partial}{\partial\mu_{q_1}}\ln\det S^{-1}(\mu_u,\mu_d,\mu_s)\right\rangle\left\langle\frac{\partial}{\partial\mu_{q_2}}\ln\det S^{-1}(\mu_u,\mu_d,\mu_s)\right\rangle \Bigg).
\eea
Therefore, the form of the fermionic determinant and how we introduce the cutoff are crucial for our calculation.

Our choice will be the following: we consider the effect of the condensates only for momenta smaller than the cutoff $\Lambda$, while in the high $p$ region only free quarks are included in the calculation, namely ($M(\sigma_\alpha,\pi_\alpha)=S^{-1}(\sigma_\alpha,\pi_\alpha)$):
\bea
	&&\sum_n\int_0^\infty\frac{\textrm{d}^3p}{(2\pi)^3}\ln\det M(\sigma_\alpha,\pi_\alpha,\omega_n)=
	\sum_n\int_0^\Lambda\frac{\textrm{d}^3p}{(2\pi)^3}\ln\det M(\sigma_\alpha,\pi_\alpha,\omega_n)\nonumber\\
	&&\hspace{2cm}+\sum_n\int_\Lambda^\infty\frac{\textrm{d}^3p}{(2\pi)^3}\ln\det M(\sigma_\alpha=0,\pi_\alpha=0,\omega_n).
\eea
For the free term of the decomposition we know the eigenvalues of the fermionic matrix analytically and therefore the sum over the Matsubara frequencies can be performed exactly. In this way the fermionic term can be rewritten as:
\bea
	&&\sum_n\int_0^\infty\frac{\textrm{d}^3p}{(2\pi)^3}\ln\det M(\sigma_\alpha,\pi_\alpha,\omega_n)=
	\sum_n\int_0^\Lambda\frac{\textrm{d}^3p}{(2\pi)^3}\ln\det M(\sigma_\alpha,\pi_\alpha,\omega_n)\nonumber\\
	 &&\hspace{2cm}+\sum_j\int_\Lambda^\infty\frac{\textrm{d}^3p}{(2\pi)^3}\ln\left[1+e^{-E_j/T}\right],
\eea
where the sum over the index $j$ means the sum over the different eigenvalues of the fermionic matrix.

\begin{center}
\begin{figure}
\begin{center}
\hspace{-3cm}
%\begin{minipage}{.48\textwidth}
%\hspace{.8cm}
\parbox{6cm}{
\scalebox{.75}{
\includegraphics{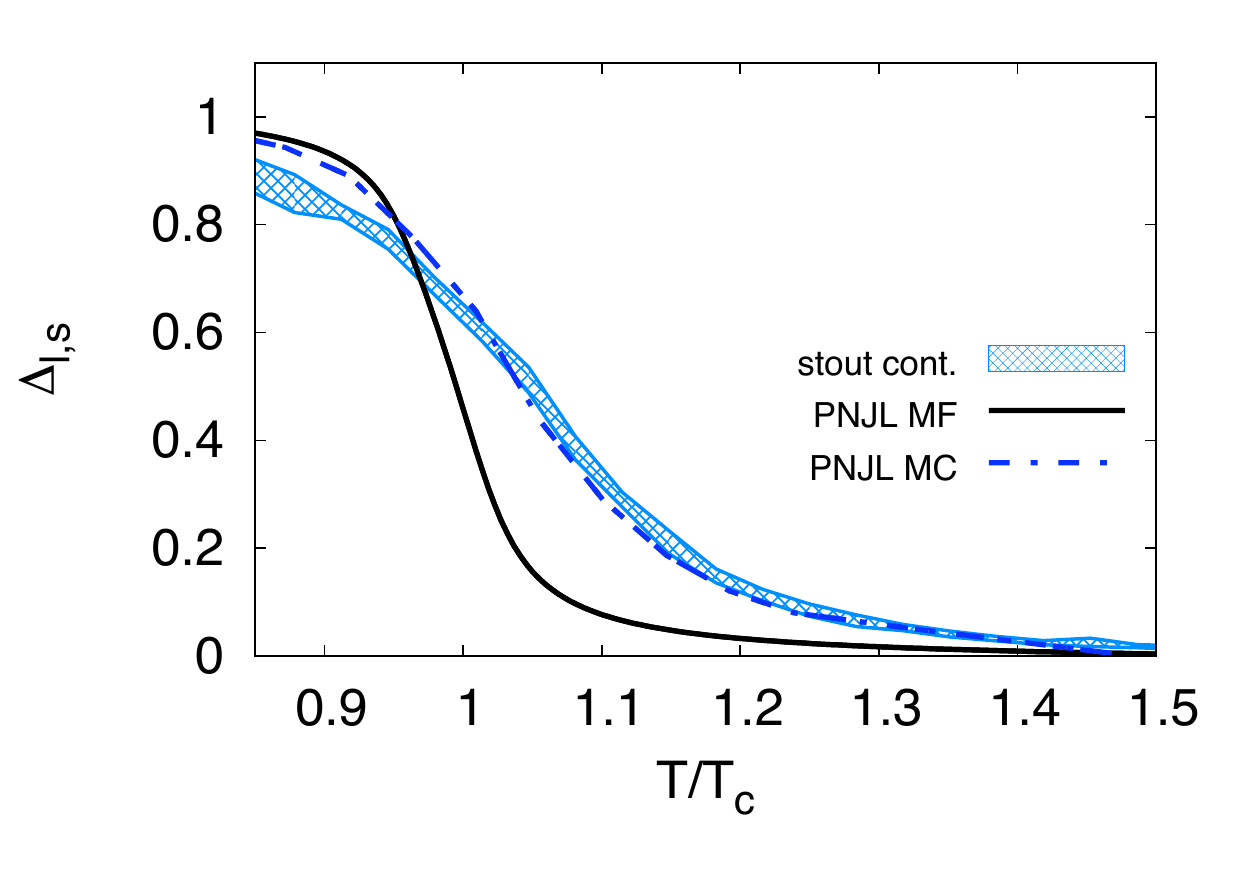}\\}}
%\end{minipage}
%\begin{minipage}{.48\textwidth}
%\hspace{.6cm}
%\parbox{6cm}{
%\scalebox{.64}{
%\includegraphics{cBS6_PNJL}\\}}
%\end{minipage}
%\centerline{(a)}
\end{center}
\caption{Renormalized chiral condensate $\Delta_{l,s}$: comparison between the lattice results (light blue band) \cite{Borsanyi:2010bp}, the mean-field PNJL model result (full, black line), and the Monte-Carlo PNJL model result (blue, dash-dotted line).  }
\label{fig3}
%\vspace{-.4cm}
\end{figure}
\end{center}

The importance of including fluctuations in the model is shown in Fig. \ref{fig3}, where we compare the chiral condensate from lattice QCD (from Ref. \cite{Borsanyi:2010bp}), the PNJL result at the mean-field level, and the PNJL result in which fluctuations of all fields are included. As it is evident, the inclusion of fluctuations makes the curve much smoother and it brings it to a good agreement with the lattice data (a similar effect was observed in Ref. \cite{Skokov:2010wb}).

\section{Results and conclusions}

In order to draw conclusions on the presence of bound states in the QGP, the most relevant comparison between PNJL and lattice results is for the non-diagonal quark correlators. In particular, we will focus here on the $u-s$ correlator and on the baryon-strangeness correlator, for which new lattice results have been recently published \cite{Borsanyi:}.  At the mean field level, the PNJL model has no correlations between the different quark flavors, therefore the corresponding $u-s$ correlator stays flat and equal to zero over the full temperature range. Similarly, the baryon-strangeness correlator takes everywhere the value corresponding to the non-interacting QGP, namely one \cite{Koch:2005vg}. In order to properly estimate
all possible contributions to this observable from colored degrees of freedom and mesonic zero modes, we need to go beyond mean field and take fluctuations of all fields (Polyakov loop, chiral condensates, pion and kaon condensates) into account.  By plotting the results corresponding to fluctuations of the Polyakov loop only, we will be able to determine the relative strength of the quark correlator due to colored states above $T_c$. The difference between such a
PNJL calculation and lattice QCD can therefore give us an estimate on the relative abundance of bound states in the medium above $T_c$. The fluctuations of the condensates give us an estimate of the contribution due to the zero-mode mesonic states. We will find that these fluctuations vanish in the infinite volume limit, leaving the Polyakov loop fluctuations as the only non-zero contribution to this observable.
\begin{figure}
\begin{minipage}{.48\textwidth}
\hspace{-.8cm}
\parbox{6cm}{
\scalebox{.66}{
\includegraphics{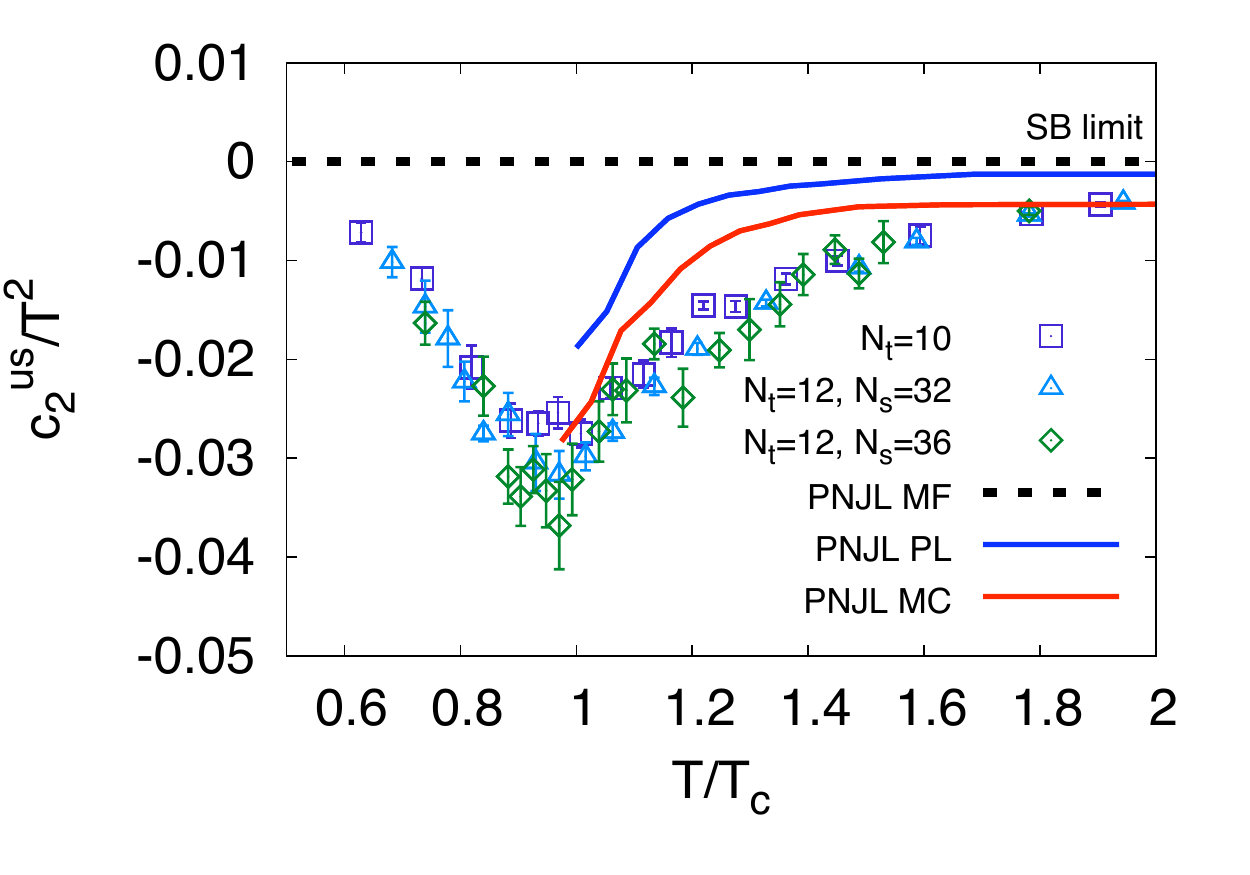}\\}}
\end{minipage}
\begin{minipage}{.48\textwidth}
%\hspace{.6cm}
\parbox{6cm}{
\scalebox{.64}{
\includegraphics{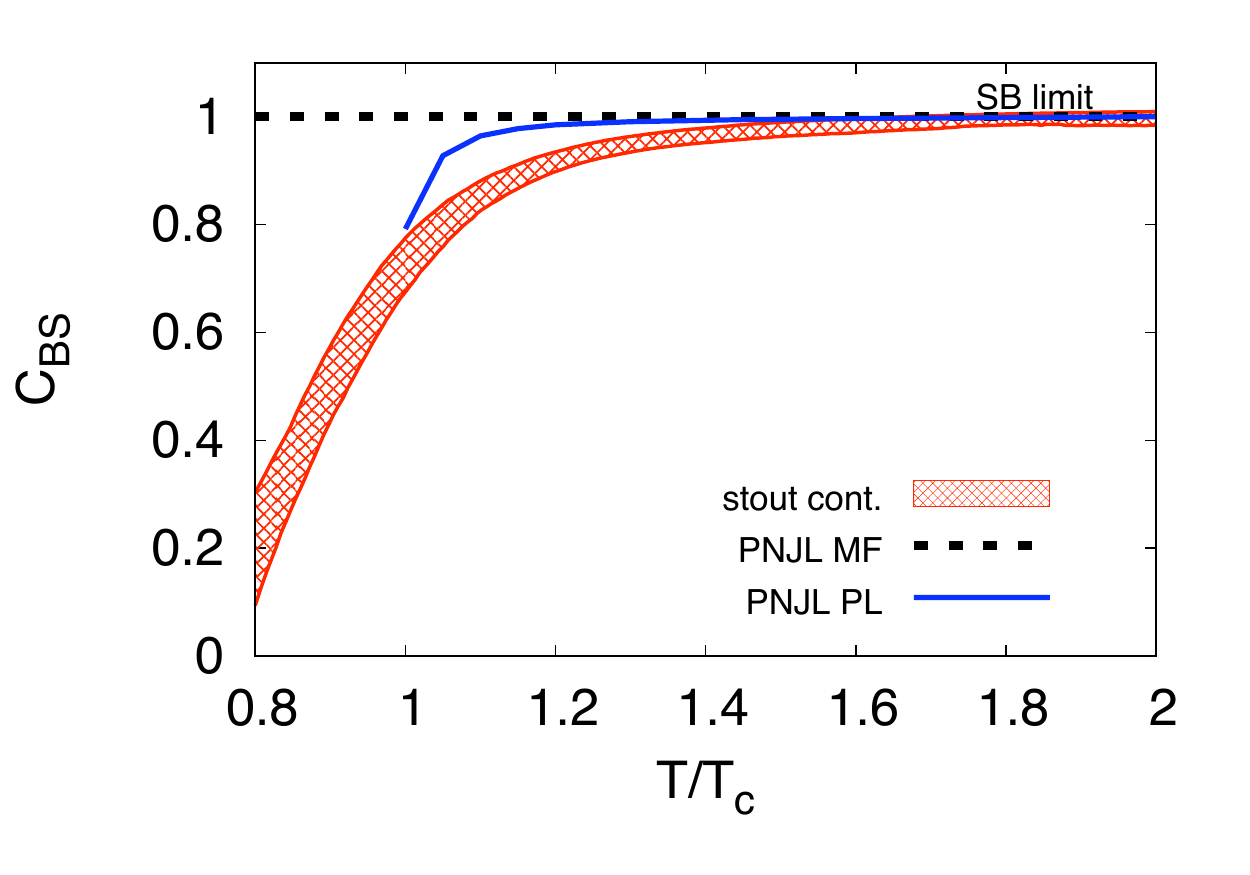}\\}}
\end{minipage}
%\centerline{(a)}
\caption{Left: Comparison between the lattice results for the $u-s$ correlator as a function of $T/T_c$ \cite{Borsanyi:}, and the PNJL model results. The mean field PNJL result is zero for all temperatures, as expected (dashed curve). The blue curve corresponds to the PNJL model result when only the Polyakov loop fluctuations are taken into account. The red curve is the full PNJL model prediction, with fluctuations of all fields taken into account. Notice that the red curve will fall on the blue curve in the infinite volume limit. Right: Baryon-strangeness correlator: comparison between the continuum extrapolated lattice results from Ref. \cite{Borsanyi:} (red band), the PNJL model result at the mean field level (black, dashed line) and the PNJL model result with inclusion of Polyakov loop fluctuations (blue, solid line).}
\label{fig2}
%\vspace{-.4cm}
\end{figure}
The results of our calculation are shown in Fig. \ref{fig2}: since the PNJL model describes the deconfined state, we only show the corresponding curves for $T>T_c$. In the left panel, the lattice results for the
$u-s$ correlator as a function of $T/T_c$ are compared to the PNJL model results at the mean field level
(dashed line), when Polyakov loop fluctuations are taken into account (blue line) and when fluctuations
of all fields are taken into account (red line). The same comparison is made in the right panel for the other observable we are considering, namely the baryon-strangeness correlator.
It is evident from the left panel, that even a PNJL model taking into account all possible fluctuations cannot fully account for the strong dip and the slow rise of the non-diagonal quark correlator determined by lattice QCD.
Notice that the red curve has been obtained in a Monte Carlo simulation for which the ratio $N_s/N_t$ (see Eq. (\ref{vol})) is the same as the one used in the lattice simulations. However, we find that in the thermodynamic limit of $N_s\rightarrow\infty$, the blue curve, which corresponds to the colored QGP contribution to this observable,
remains the same, while the red curve falls on top of the blue curve, i.e. fluctuations of the condensates (zero-mode mesonic contributions) vanish in the infinite volume limit. It is evident that the lattice data can be reproduced only for large temperatures, thus allowing for a considerable contribution from finite-momentum bound states of both baryonic and mesonic nature.

In the right panel we show the baryon-strangeness correlator. A comparison between lattice data and PNJL model results for this observable again suggests the presence of bound states in the QGP for temperatures up to 1.6-1.7 $T_c$.

\begin{center}
\begin{figure}
\begin{center}
\hspace{-5cm}
%\begin{minipage}{.48\textwidth}
%\hspace{.8cm}
\parbox{6cm}{
\scalebox{.12}{
\includegraphics{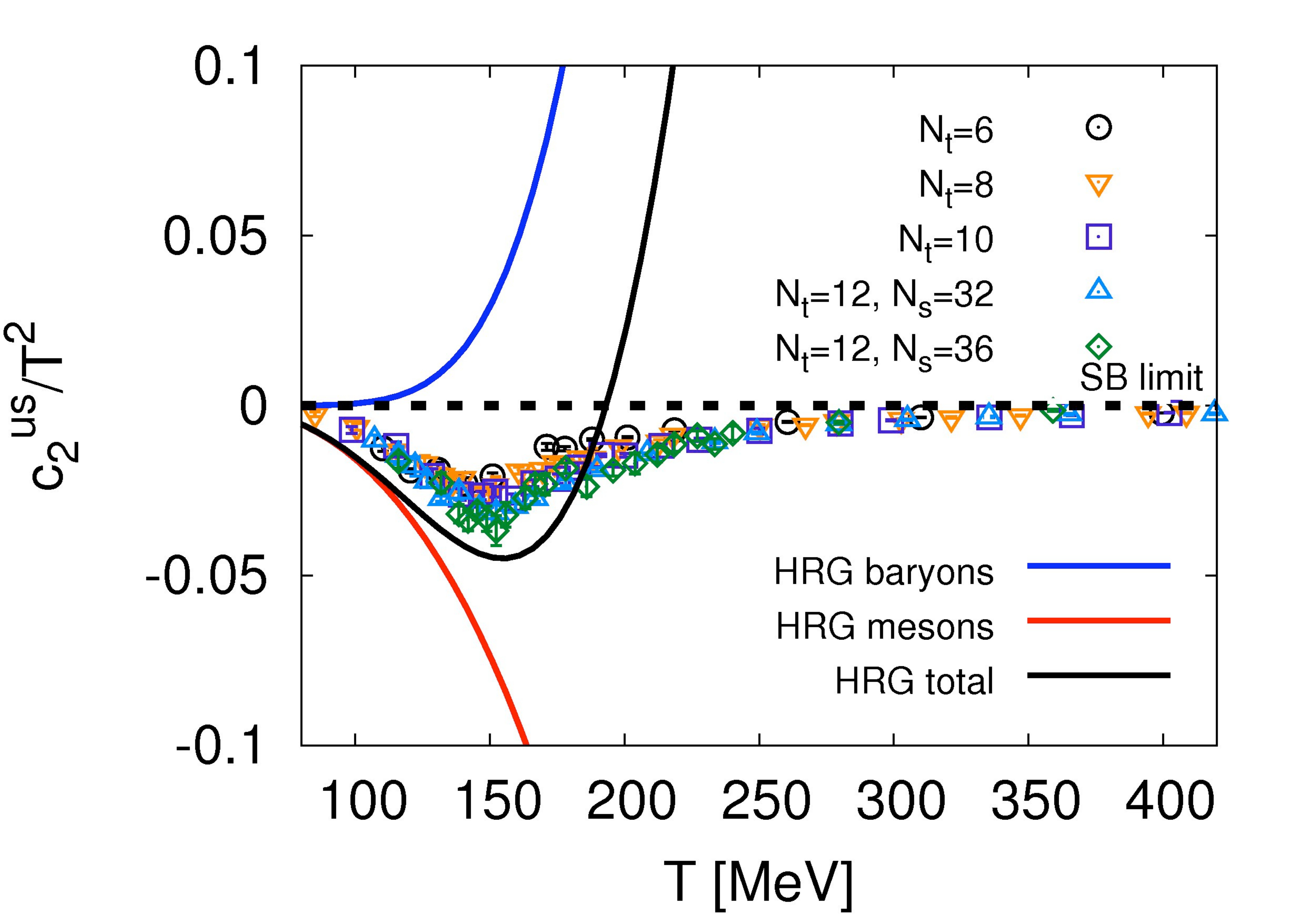}\\}}
%\end{minipage}
%\begin{minipage}{.48\textwidth}
%\hspace{.6cm}
%\parbox{6cm}{
%\scalebox{.64}{
%\includegraphics{cBS6_PNJL}\\}}
%\end{minipage}
%\centerline{(a)}
\end{center}
\caption{$u-s$ correlator: the different curves show the contributions of baryons (blue), mesons (red) and the total (black) from the Hadron Resonance Gas model. }
\label{fig4}
%\vspace{-.4cm}
\end{figure}
\end{center}

Furthermore, the peculiar shape of the non-diagonal correlator near $T_c$ (sharp dip and subsequent slow rise towards zero) can be interpreted when comparing it to the separate contributions of mesonic and
baryonic states in a hadron resonance gas (HRG) calculation \cite{hrg}.
Mesonic states in the HRG model exhibit a negative correlation, whereas baryonic states yield a positive value (see Fig. \ref{fig4}). The dip and slow rise thus is likely caused by enhanced baryonic state formation (or survival) at higher temperatures. Still, the lattice data never exceed zero before full deconfinement is reached, which means that baryonic states never dominate the hadron formation. This is confirmed by the magnitude of the difference between PNJL and lattice QCD in C$_{BS}$.

Taking into account the flavor dependence of the light quark susceptibilities as shown in the right
panel of Fig. \ref{fig1}, one can deduce a scenario where strange quark bound states are formed (or survive) at higher $T$ in the deconfined medium than
light quark bound states. Earlier lattice calculations only exhibited this effect in the comparison between light and heavy quarks, but the recent improvement in lattice accuracy indicates effects already at the
strange quark level, which leads to specific experimentally verifiable predictions, such as an enhanced
survival probability of strange over non-strange resonances near, but above, $T_c$.

In summary, a comparison of PNJL and lattice QCD calculations yields ample evidence, that a phase of mixed degrees of freedom exists for a particular temperature range above the QCD transition temperature. The PNJL model calculations and the mapping of the flavor and baryon number dependence in lattice QCD calculations, show that it is likely that in this equilibrated phase the relative abundance of strange hadronic and baryonic states is larger at high $T$ rather than immediately above $T_c$, but mesonic bound states still dominate over the entire temperature range. In other words, strange hadrons and non-strange baryons form earlier in the mixed phase than
their non-strange and mesonic partners. These dependencies should be experimentally verifiable through particle identified measurements of light- and strange-quark hadrons and resonances near the phase boundary.

\section*{Acknowledgements}
We acknowledge fruitful discussions with V. Koch and K. Redlich. R. B. thanks the Department of Experimental Physics of Torino University for the hospitality during the period when this work was developed. His work is supported by the U.S Department of Energy under DOE grant: DE-FG02-07ER41521. The work of C. R. is supported by funds provided by the Italian Ministry of Education, Universities and Research under the Firb Research Grant
RBFR0814TT. The Monte Carlo PNJL calculations were performed using the Aurora supercomputer at FBK/Trento.  The work of M. C. is supported by the AuroraScience project, which is funded jointly by the Provincia Autonoma di Trento (PAT)  and the Istituto Nazionale di Fisica Nucleare (INFN).

\end{document}